\documentclass[twocolumn,aps,pra,amsmath,amssymb,amsfonts,superscriptaddress,notitlepage,reprint,longbibliography,floatfix
]{revtex4-1}
\usepackage{graphicx}
\usepackage{dcolumn}
\usepackage{bm}
\usepackage{amssymb}
\usepackage[svgnames]{xcolor}
\usepackage[colorlinks=true,
            linkcolor=red,
            urlcolor=blue,
            citecolor=DarkGreen]{hyperref}

\hypersetup{
    pdftitle = {Realizing a Deterministic Source of Multipartite-Entangled Photonic Qubits},
    pdfauthor = {Jean-Claude Besse}
}

\usepackage[english]{babel}
\usepackage{braket}

\usepackage{microtype}
\usepackage{bbm}

\begin{document}

\title{Realizing a Deterministic Source of Multipartite-Entangled Photonic Qubits}

\author{Jean-Claude~Besse}\email{jbesse@phys.ethz.ch}
\author{Kevin Reuer}
\author{Michele C. Collodo}

\author{Arne Wulff}
\author{Lucien Wernli}
\author{Adrian Copetudo}
\affiliation{Department of Physics, ETH Zurich, CH-8093 Zurich, Switzerland}

\author{Daniel Malz}
\affiliation{Max-Planck-Institute of Quantum Optics, Hans-Kopfermann-Str.~1, 85748 Garching, Germany}
\affiliation{Munich Center for Quantum Science and Technology, Schellingstr.~4, 80799 M\"{u}nchen, Germany}

\author{Paul Magnard}
\author{Abdulkadir Akin}
\author{Mihai Gabureac}
\author{Graham J. Norris}
\affiliation{Department of Physics, ETH Zurich, CH-8093 Zurich, Switzerland}

\author{J. Ignacio Cirac}
\affiliation{Max-Planck-Institute of Quantum Optics, Hans-Kopfermann-Str.~1, 85748 Garching, Germany}
\affiliation{Munich Center for Quantum Science and Technology, Schellingstr.~4, 80799 M\"{u}nchen, Germany}

\author{Andreas Wallraff}
\author{Christopher Eichler}\email{eichlerc@phys.ethz.ch}
\affiliation{Department of Physics, ETH Zurich, CH-8093 Zurich, Switzerland}

\date{\today}

\begin{abstract}
Sources of entangled electromagnetic radiation are a cornerstone in quantum information processing and offer unique opportunities for the study of quantum many-body physics in a controlled experimental setting. While multi-mode entangled states of radiation have been generated in various platforms, all previous experiments are either probabilistic or restricted to generate specific types of states with a moderate entanglement length. Here, we demonstrate the fully deterministic generation of purely photonic entangled states such as the cluster, GHZ, and W state by sequentially emitting microwave photons from a controlled auxiliary system into a waveguide. We tomographically reconstruct the entire quantum many-body state for up to $N=4$ photonic modes and infer the quantum state for even larger $N$ from process tomography. We estimate that localizable entanglement persists over a distance of approximately ten photonic qubits, outperforming any previous deterministic scheme.
\end{abstract}

\maketitle

Entanglement is one of the most fundamental concepts in quantum physics~\cite{Guhne2009} and an essential resource for applications in quantum information processing~\cite{Kempe1999,Zang2015}. Both the theory of entanglement~\cite{Horodecki2009} and the experimental generation of entangled states of light~\cite{Eibl2004,Walther2005,Wang2018a,Schwartz2016,Istrati2019,Takeda2019} and matter~\cite{Friis2018,Omran2019a,Wei2020} have therefore been subject of intense research. Of particular importance are multi-partite entangled states of photons for their use in quantum communication and network protocols~\cite{Gisin2007}. Experiments to generate entangled states of light most commonly rely on spontaneous parametric down-conversion sources and heralding~\cite{Eibl2004,Walther2005,Wang2018a}. The probabilistic nature of such schemes is a major obstacle when scaling to larger systems, which has motivated the study of deterministic sources of entangled photonic states more recently~\cite{Schwartz2016,Istrati2019,Takeda2019}. So far, only states within device-specific classes of entanglement have been generated deterministically and with a moderate size as compared to their matter-based counterparts~\cite{Friis2018,Wei2020}. Achieving more versatility in the generation of entanglement has therefore been an outstanding challenge which we address in this work.

\begin{figure}[!t]
\includegraphics[width=\columnwidth]{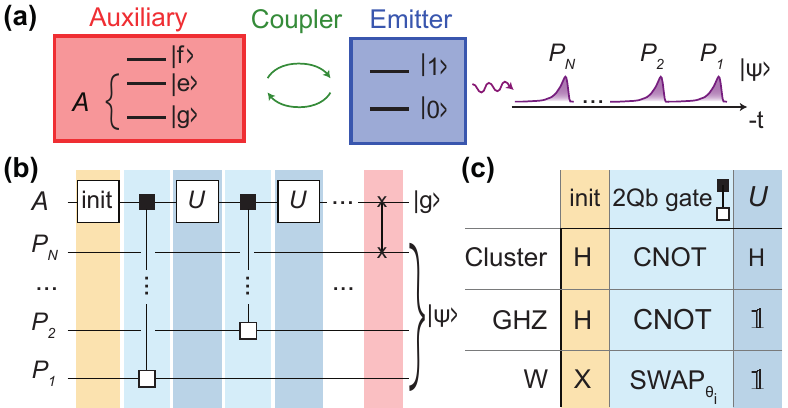}
\caption{\label{fig:circuit_and_timetraces}\textbf{Generation of entangled states of photons.} (a) An auxiliary quantum system $A$ coupled to an emitter is used to generate a state $\ket{\psi}$ of photonic modes $P_i$.
(b) Quantum circuit using an initialization gate and single-qubit gates $U$ acting on the auxiliary system $A$, and two-qubit gates (small squares) of the CNOT and SWAP family between the auxiliary qubit and the photons $P_i$. The last two-qubit gate is always a SWAP (crosses).
(c) Specific gates used for the generation of the cluster state, the GHZ state, and the W state. \textsf{H}: Hadamard gate, $\mathbbm{1}$: Identity gate.
}
\end{figure}
\begin{figure*}[!t]
\includegraphics[width=\textwidth]{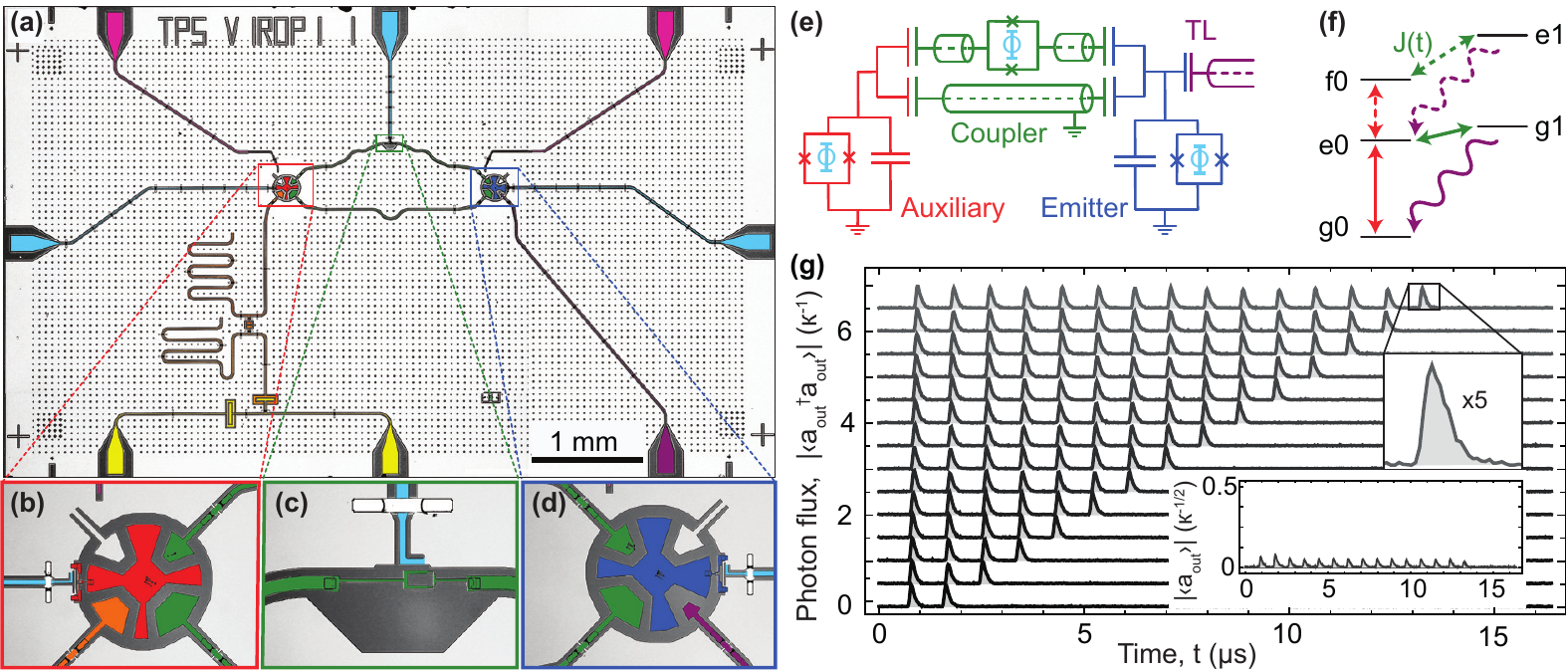}
\caption{\label{fig:experimental_setup}\textbf{Experimental cQED implementation.} (a) False colour optical micrograph of the chip used in the experiment, comprising an auxiliary qutrit (red, b), a tunable coupler (green, c), an emitter qubit (blue, d), an output transmission line (purple) connecting to the measurement chain, a feedline (yellow), a readout circuit with Purcell filter (orange), three flux lines (cyan) and two charge lines (pink). (e) Equivalent electrical circuit diagram. The auxiliary and the emitter qubits are transmons, with a flux tunable superconducting quantum interference device (SQUID) loop capacitively shunted to ground. The coupler consists of two paths, one mainly inductive (top), the other capacitive (bottom). Photons are emitted into the semi-infinite transmission line (TL).  (f) Energy level diagram, indicating the possible transitions induced by single qubit gates (red), parametric modulation of the coupler (green), and photon emission via spontaneous decay into the TL (purple).
(g) Measured outgoing photon flux $\left|\langle a_\text{out}^\dagger a_\text{out} \rangle \right|$ versus time in the transmission line, in units of inverse emitter linewidth $\kappa$, for cluster states of $N=2,...,15$ modes. Traces are offset by 0.5 for clarity. Insets: zoom-in by a factor 5 of the last time-bin for clarity, and amplitude $\left|\langle a_\text{out} \rangle\right|$ in units of $\kappa^{-1/2}$, for the cluster state of 15 modes.}
\end{figure*}

A generic protocol to generate entangled states as a train of sequentially emitted photons was proposed by Sch\"{o}n et al.~\cite{Schoen2005a} and is based on a long-lived auxiliary quantum system $A$, which sequentially interacts with an emitter qubit via a controllable coupling, see Fig.~\ref{fig:circuit_and_timetraces}(a). After each interaction cycle $i$ the engineered emission into a waveguide converts the state of the emitter qubit into a flying photonic qubit $P_i$ defined by the presence or absence of a single photon in the associated time bin. The combination of controllable interaction and photon emission can be understood as a two-qubit gate realized between $A$ and $P_i$. Consequently, the photonic state preparation is formally described by repeated unitary operations $U$ applied to $A$ interleaved with two-qubit gates, as represented by the generic quantum circuit shown in Fig.~\ref{fig:circuit_and_timetraces}(b). The range of accessible states crucially depends on the set of available two-qubit gates. While the GHZ state and the cluster state require a controlled NOT (CNOT) gate only~\cite{Lindner2009} and have already been generated experimentally~\cite{Schwartz2016}, the preparation of a W state relies on the ability to perform SWAP($\theta_i$) gates with adjustable rotation angle $\theta_i=2\arcsin \left[(N-i+1)^{-0.5}\right]$, see Fig.~\ref{fig:circuit_and_timetraces}(c). Furthermore, a final SWAP operation is required to disentangle the photonic qubits from the auxiliary system by bringing the latter back to its ground state $\ket{g}$, thereby rendering the generation of the final photonic state $|\psi\rangle$ fully deterministic.

In this work, we realize a superconducting circuit-based source of entangled microwave photons (see Fig.~\ref{fig:experimental_setup}) and implement the above protocol with both generic SWAP- and CNOT-type gates to demonstrate deterministic state  preparation of cluster, GHZ and W states. We realize the auxiliary system as a transmon (red) of which we use the first three energy level $\ket{g}$, $\ket{e}$, and $\ket{f}$ with transition frequencies $\omega_{ge}/2\pi= 5.758~$GHz and $\omega_{ef}/2\pi= 5.455~$GHz, see App.~\ref{SM:cabling} and \ref{SM:characterization} for details about the device fabrication, the experimental setup, and sample characterization. We perform local unitary operations between the states of the auxiliary qutrit system by applying microwave pulses with controlled amplitude and phase resonant with $\omega_{ge}$ and $\omega_{ef}$, respectively [red arrows in Fig.~\ref{fig:experimental_setup}(f)].
In order to stimulate state-dependent photon emission processes, we tunably couple the auxiliary transmon to a second transmon (blue), acting as the emitter, which is operated at $\omega_{01}/2\pi= 5.896~$GHz and decays into a semi-infinite transmission line with rate $\kappa/2\pi = 1.95~$MHz.
The coupling between the auxiliary system and the emitter is mediated by two parallel channels, one of which is tunable via the magnetic flux applied to a superconducting quantum interference device (SQUID) loop. This specific coupler arrangement (green) allows us to interferometrically cancel the static coupling and protect the auxiliary transmon from Purcell decay into the transmission line while enabling fast decay of the emitter~\cite{Collodo2019,Mundada2019}.

By parametrically modulating the applied flux around the value at which the static coupling is cancelled, we selectively drive sideband transitions between excitation-number-conserving states [green arrows in Fig.~\ref{fig:experimental_setup}(f)] with a rate $J_\text{ac}/2\pi \simeq 5~$MHz. To perform SWAP($\theta_i$) gates between the auxiliary qubit and the emitter qubit, we modulate the flux at the difference frequency $\Delta_\text{e0g1}/2\pi=(\omega_{01}-\omega_{ge})/2\pi= 139~$MHz driving a transition between the states $\ket{e0}$ and $\ket{g1}$. The CNOT gate is a photon emission process conditioned on the auxiliary qubit being in its first excited state. We realize this state-dependent photon generation by first applying a $\text{R}_\text{y}(\pi)$ pulse on the $e$-$f$ transition of the auxiliary qubit and then driving a sideband transition at frequency $\Delta_\text{f0e1}/2\pi=(\omega_{01}-\omega_{ef})/2\pi=441~$MHz between the $\ket{f0}$ and $\ket{e1}$ state, which brings the auxiliary qubit back into the $\ket{e}$ state while emitting a photon.

By controlling amplitude, duration and phase of the pulses we choose any targeted SWAP and CNOT angle~\cite{Abrams2019a,Foxen2020}, see App.~\ref{SM:2qbgates} for details. Using a continuous set of two-qubit gates allows for the generation of a wide family of entangled states belonging to the class of matrix-product states (MPS) with bond dimension $d=2$~\cite{Cirac2010a}.

\begin{figure}[b]\setlength{\hfuzz}{1.1\columnwidth}
\begin{minipage}{\textwidth}
\includegraphics[width=\textwidth]{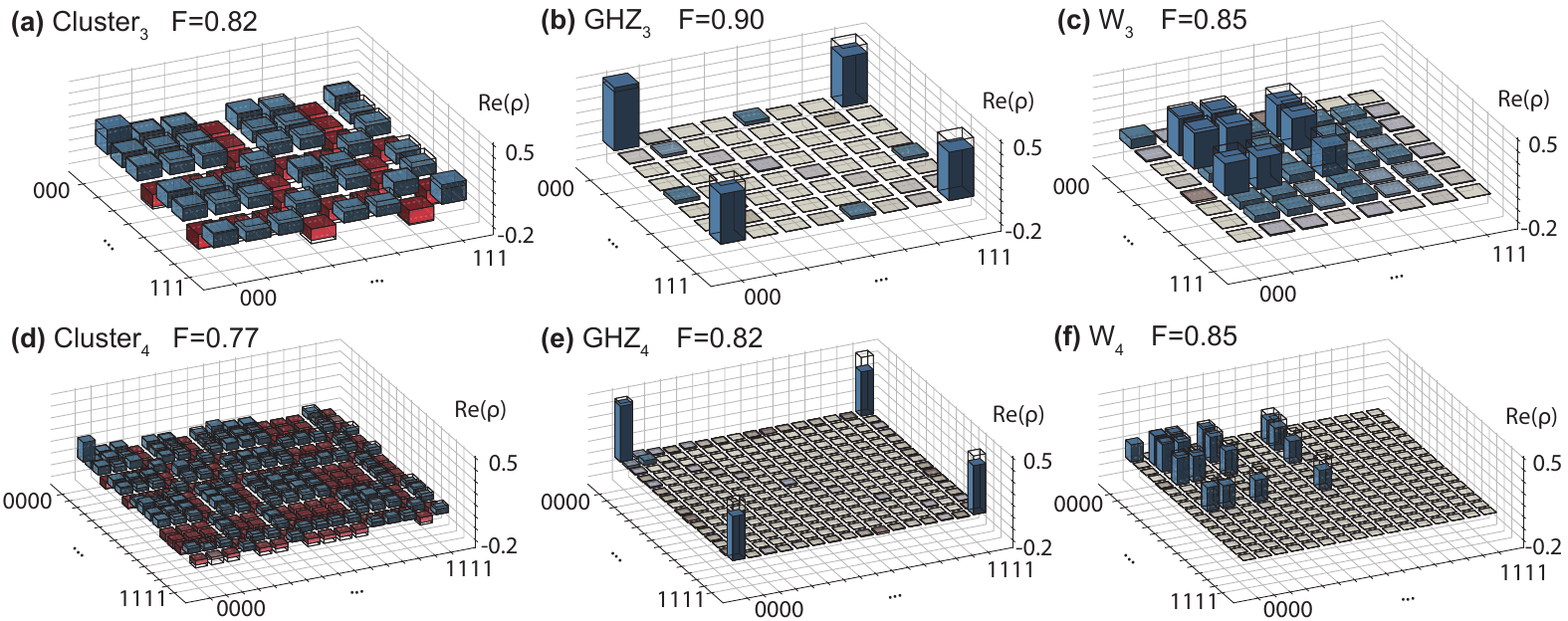}
\caption{\label{fig:3and4photons}\textbf{Complete state tomography.} (a-c) 3- and (d-f) 4-mode state tomography. We plot the real part $\text{Re} (\rho)$ of the most likely density matrices $\rho$ (bars) for (a,d) the cluster state, (b,e) the GHZ state, and (c,f) the W state. Absolute values of the imaginary part (not shown) are all below 0.03. Ideal density matrices $\rho_\text{ideal}$ are shown as black wireframes. The fidelity $F=\text{Tr}\left( \sqrt{\sqrt{\rho}\rho_\text{ideal}\sqrt{\rho}} \right)^2$ is indicated. Entries are following the binary ordering from (0)000 to (1)111.}
\end{minipage}
\end{figure}
By applying sequences of the described gates, according to the quantum circuit shown in Fig.~\ref{fig:circuit_and_timetraces}(b-c), we generate many-body states of microwave radiation and experimentally assess their properties through heterodyne measurements of the output field $a_\text{out}$ in the transmission line, see App.~\ref{SM:tomography} for information about the detection method. We first measure the temporal profile of the pulse train by averaging the photon flux $\left|\langle a_\text{out}^\dagger a_\text{out} \rangle\right|$ over $10^6$ repetitions of the experiment, see Fig.~\ref{fig:experimental_setup}(g) for the example of prepared cluster states with up to 15 photonic qubits. Each pulse has an initial rise time of approximately $\pi/J_{\rm ac}\simeq100\,$ns corresponding to the duration of the two-qubit gate, before exponentially decaying with characteristic timescale $\kappa^{-1}\simeq 80$~ns given by the decay rate of the emitter into the transmission line (see inset). We have chosen a conservative repetition time of $T = 900$~ns, to ensure that the emitter has fully returned into its ground state before starting the next emission cycle. Each time bin contains an integrated photon flux of ideally half a photon for the cluster state. In comparison to the photon flux, the field expectation value $\left|\langle a_\text{out} \rangle \right|$, shown in the inset, is close to zero, as expected for the cluster state. We attribute the small but finite measured $\left|\langle a_\text{out} \rangle\right|$ to coherent errors at the 1\% level in single-qubit $e$-$f$ pulses.

To extract the quantum-mechanical properties of the generated many-body states we perform quantum state tomography. We reconstruct the density matrix for up to four modes, enabled by advancing the capabilities of our field programmable gate array (FPGA)-based data acquisition, see Fig.~\ref{fig:3and4photons} (App.~\ref{SM:tomography} for a detailed discussion of the tomography method).
Reconstructing the joint density matrix of four modes is well beyond what has been demonstrated previously for propagating microwave fields, which is full tomography for up to two modes~\cite{Lang2013,Kannan2020}. We note that full tomography for even more than four modes is possible by further extending the capacity of data storage on the FPGA or by using offline data processing solutions.

{
We find that the cluster state contains non-vanishing values in all entries of the density matrix, with magnitude close to $2^{-N}$, indicating that all basis states are occupied with nearly equal probability. The characteristic pattern of sign changes in some individual terms, which cannot be produced by local $Z$ gates as there is an imbalance between the number of basis states with positive and negative sign, renders this state inseparable. For the GHZ state, defined as the equal superposition of the absence or presence of a single photon in each mode $\text{GHZ}_N=\left(\ket{0}^{\otimes N}+\ket{1}^{\otimes N}\right)/\sqrt{2}$, only the four corners of the density matrix have a non-zero value, equal to 0.5. Finally, the W state $\text{W}_N=\sum_{i=1}^N \ket{0}^{\otimes (i-1)}\ket{1_i}\ket{0}^{\otimes (N-i)}/\sqrt{N}$ is restricted to entries containing exactly one excitation. All non-vanishing entries have ideally the same value of $N^{-1}$. In all six cases the experimentally reconstructed density matrices are in very good agreement with the ideally expected ones, which is indicated by the high fidelities $F_\text{Cluster}=0.82\,(0.77)$, $F_\text{GHZ}=0.90\,(0.82)$, and $F_\text{W}=0.85\,(0.85)$ for $N=3\,(4)$, with $F=\text{Tr}\left( \sqrt{\sqrt{\rho}\rho_\text{ideal}\sqrt{\rho}} \right)^2$, compared to the respective ideal state $\rho_\text{ideal}$. For nearly all prepared states we observe a higher probability to be in the ground state compared to the ideal case and an overall decrease in the population of states with photons emitted in late time bins. We attribute these effects to the relaxation of the auxiliary qubit during the sequential emission process. Indeed, master equation simulations, taking finite relaxation times $T_1^{(e)}=21~\mu$s ($T_1^{(f)}=7~\mu$s) \unskip\parfillskip 0pt \par}
\newpage
\noindent and Ramsey dephasing times $T_2^{\star(e)}=17~\mu$s ($T_2^{\star(f)}=8~\mu$s) into account, indicate that the majority of the infidelity originates from the finite coherence of the auxiliary qubit. For the 4-mode states for example, we simulate fidelities $F_\text{Cluster}^\text{(MES)}=0.87$, $F_\text{GHZ}^\text{(MES)}=0.89$, and $F_\text{W}^\text{(MES)}=0.90$ to the respective ideal state. In general, the W state is affected by the decoherence of the first-excited level only, but requires the experimental calibration of more gates. The GHZ state has a higher simulated fidelity at large mode number $N$ since its overlap with the vacuum state is 0.5, whereas this overlap scales at $1/N$ for the cluster state.

\begin{figure}[!t]
\includegraphics[width=\columnwidth]{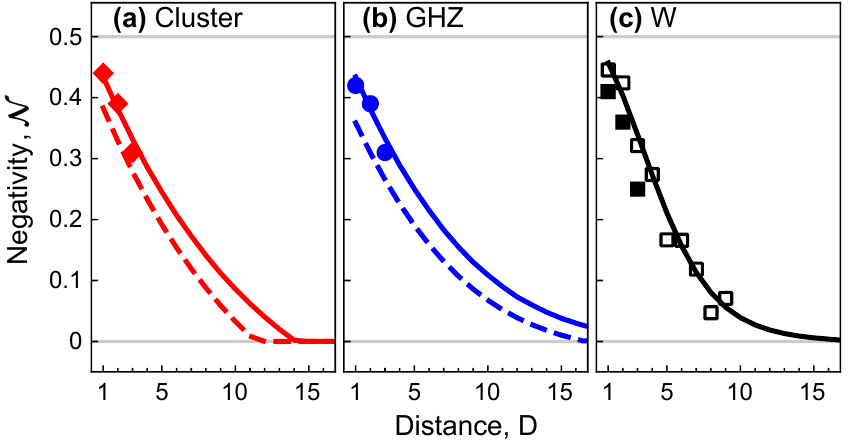}
\caption{\label{fig:negativity-maps}\textbf{Localizable entanglement.} Negativity $\mathcal{N}$ between the first and $(D+1)$-th photonic mode separated by a distance $D$. Filled dots indicate measurements based on complete tomography, dashed lines are inferred from process maps, empty dots are calculated from partial tomography (see text), and solid lines indicate the coherence limit from master equation simulations. Ideal value of $\mathcal{N}$ is 0.5. The cluster state (a, red diamond), GHZ state (b, blue circles), and W state (c, black squares) data are shown.}
\end{figure}

To quantify entanglement in the generated many-body states of light, we analyze the localizable entanglement between the first and the last qubit in the chain~\cite{Verstraete2004} after projecting out all other qubits to obtain the two-qubit density matrix $\rho_D$ as a function of their distance $D=N-1$.
As a metric, we choose to report the negativity $\mathcal{N}(\rho_D)=\sum_i \left| \lambda_i \right|$, where $\lambda_i$ are the negative eigenvalues of the partial transpose of $\rho_D$, which is an entanglement monotone for bipartite systems and hence a suited measure of entanglement. For the data points obtained from full tomography (see Fig.~\ref{fig:negativity-maps}, filled markers) $\rho_D$ is calculated from the full density matrix by projecting all qubits but the first and last into their respective ground states~\footnote{The original definition of localizable entanglement averages over all possible outcomes in a given measurement basis. For the W state in particular, localizable entanglement only persists in case the measurement outcome is the ground state for all measured qubits. The resulting estimate for the entanglement length is independent of this particular choice.} in the $X$- ($Z$-)basis for the cluster and GHZ (W) states.
The resulting negativities $\mathcal{N}_\text{Cluster}=0.44,0.39,0.31$, $\mathcal{N}_\text{GHZ}=0.42,0.39,0.31$, and $\mathcal{N}_\text{W}=0.41,0.36,0.25$ decrease monotonically with $N=2,3,4$   ($D=1,2,3$) as expected and in good agreement with master equation simulations (solid lines) taking the finite coherence of the auxiliary system into account.

While tomography of the most general $N$-qubit state requires exponentially growing resources for data acquisition and processing, MPS are fully characterized by the process maps applied in each emission cycle to the auxiliary qubit and the respective photonic mode. In order to estimate the maximal size of states over which entanglement persists, we thus infer the entanglement length $D_\text{ent.}$ from measured process maps for the cluster and GHZ states and from partial tomography for the W state as detailed in the following.
We characterize the generated cluster and GHZ states for $N>4$ considering the repetitive nature of the underlying photon emission processes. We tomographically measure the process map $\chi: \rho_A^{(\text{pre})} \rightarrow \rho_{A,P}^{(\text{post})}$ taking a generic input density matrix of the auxiliary qubit $\rho_A^{(\text{pre})}$ to its corresponding output state while generating an entangled photonic qubit $\rho_{A,P}^{(\text{post})}$. This measured process map, which we assume to be the same for all $N-1$ emission steps except for the last one, allows us to infer the most likely density matrix within the class of matrix product density operators with bond dimension $d=2$~\cite{Schwartz2016}, see App.~\ref{SM:maps}.
We determine the negativity between the first and last photonic qubits from the density matrices calculated on the basis of the measured process maps (dashed lines), after projecting out all remaining qubits as before. We find a larger entanglement length $D_\text{ent.,GHZ}\simeq 16$ for the GHZ state, compared to the entanglement length $D_\text{ent.,Cluster}\simeq 11$ of the cluster state, in agreement with master equation simulations.
The negativities obtained from measured process maps follow a similar trend as the ones obtained from master equation simulations, indicating that the experimental performance is mostly limited by decoherence of the auxiliary system, relative to the repetition time $T$.
The slightly lower negativity for the process map approach is likely dominated by coherent errors in the gate operations.

For the W state, each emission step is a SWAP($\theta_i$) gate with $\theta_i$ being different in each cycle $i$. Instead of measuring $N$ different process maps, we opted for an alternative characterization method, by measuring the density matrices $\rho_{\text{mix.},i}$ between pairs of the first and the $i$-th photonic qubit. For an ideal $N$-qubit W state, we expect these density matrices to satisfy the relation $\rho_{i} =  N/2\left[\rho_{\text{mix.},i}- (N-2)/N\ket{00}\bra{00}\right]$ with $\rho_i =\ket{\text{W}_2}\bra{\text{W}_2}$ independent of $i$. The second term on the right originates from tracing out $N-2$ photonic qubits.
Motivated by these identities, we calculate the experimental $\rho_i$ based on the measured $\rho_{\text{mix.},i}$ using the above relation for the W state of 10 photonic modes. We estimate the
degree of entanglement between the first and $i$-th photonic qubit from the negativity of $\rho_i$ (empty dots). Also in this case, we find good agreement with the simulation results and an entanglement length on the order of ten.

As interesting future directions one could explore applications of the presented source for one-way quantum computing with cluster states~\cite{Raussendorf2001,Walther2005,Nielsen2005}, Heisenberg-limited metrology or teleportation with GHZ states~\cite{Greenberger1990,Lee2019a,Peniakov2020}, or photon loss resilient quantum communication with W states~\cite{Dur2001,Kurpiers2019}. In addition, this versatile source of quantum many-body states of electromagnetic radiation, able to perform generic gates of the CNOT and SWAP families, could be used to access a larger variety of quantum many-body states in the MPS family, for example as ground states of variational quantum algorithms~\cite{Eichler2015,Smith2019}. Finally, we note that our platform naturally allows for integration of additional auxiliary and emitter qubits, suggesting a path to explore entangled tensor network states in higher dimensions~\cite{Gimeno-Segovia2019}.

\section*{Data availability statement}
The data produced in this work is available from the corresponding authors upon reasonable request.

\section*{Acknowledgments}
The authors thank Will Oliver and Barath Kannan for comments on the manuscript.
This work was supported by the Swiss National Science Foundation (SNSF) through the project ``Quantum Photonics with Microwaves in Superconducting Circuits'', by the European Research Council (ERC) through the project ``Superconducting Quantum Networks'' (SuperQuNet), by the National Centre of Competence in Research ``Quantum Science and Technology'' (NCCR QSIT), a research instrument of the Swiss National Science Foundation (SNSF), and by ETH Zurich. J.I.C. acknowledges funding from the ERC Grant QUENOCOBA.

\section*{Author Contributions}
J.-C.B. and K.R. designed the device. J.-C.B., G.N. and M.G. fabricated the device. K.R., P.M. and K.A. programmed the experimental data acquisition device. \mbox{J.-C.B.,} K.R. and M.C. prepared the experimental setup. J.-C.B. and K.R. characterized and calibrated the device and the experimental setup. P.M. implemented the qubit reset. J.-C.B., K.R., A.Wu., L.W. and A.C. carried out the experiments and analyzed the data, with theory support from D.M. and J.I.C.. J.-C.B. prepared the figures. J.-C.B. and C.E. wrote the manuscript with input from all co-authors. A.Wa. and C.E. supervised the work.

\section*{Competing interests}
The authors declare no competing interests.

\appendix
\section*{Supplementary Information}

\section{Fabrication and experimental setup}\label{SM:cabling}
\begin{figure*}[!t]
\includegraphics[width=\textwidth]{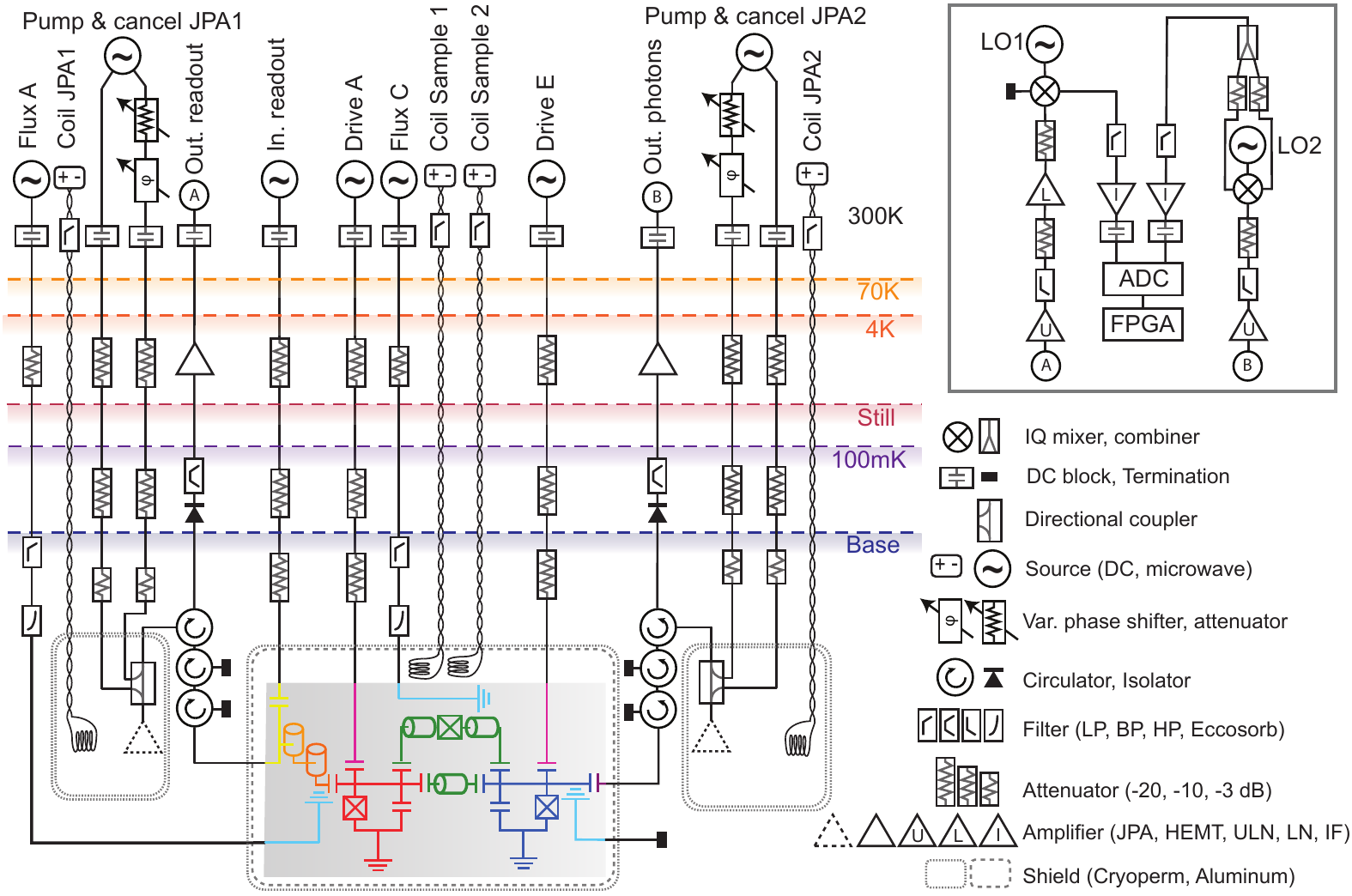}
\caption{\label{fig:fridge_setup}\textbf{Schematic of experimental setup.} For details see main text.}
\end{figure*}

We fabricated the sample, shown in Fig.~\ref{fig:experimental_setup}(a-d), on a 4.3$\,$mm x 7$\,$mm silicon substrate. We patterned all elements except for the Josephson junctions in a 150$\,$nm-thick sputtered niobium film using photolithography and reactive ion etching. We fabricated the Josephson junctions in a separate step using electron-beam lithography and shadow-evaporation of aluminum in an electron-beam evaporator.
We mounted the sample on the base temperature stage (20$\,$mK) of a dilution refrigerator, inside an aluminum and a cryoperm shield, as shown in the wiring diagram in Fig.~\ref{fig:fridge_setup}.
We used a combination of coils, and flux lines equipped with low pass (LP, 780~MHz) and Eccosorb filters, for flux biasing of the sample. We attenuated the signals in the drive lines using a standard 20/20/20~dB scheme (at the 4~K, 100~mK, and base temperature stages). We operated both the readout output line, and the photon output with Josephson Parametric Amplifiers (JPA) as first elements in the detection chains, with 27~dB and 18~dB of gain respectively. We placed them in separate sample holders with two layers of cryoperm shielding and a coil for magnetic flux biasing. At room temperature, we amplified the readout output (A) with one extra stage compared to the photon output (B). We made this choice to preserve the linearity in the amplification chain for the photonic modes, an essential requirement~\cite{Eichler2012} to perform tomography. The use of an IQ combiner increases the efficiency of the detection chain when the gain of the JPA is not fully overcoming the noise added later in the chain.

\section{Device characterization}\label{SM:characterization}
We characterize first the auxiliary system using its dispersive coupling to the readout resonator and Purcell filter~\cite{Walter2017}. Transmission of a weak signal through the feedline shows the presence of a resonator mode at frequency $\omega_\text{ro.}/2\pi=6.664$~GHz, with effective linewidth $\kappa_\text{eff.}/2\pi=25$~MHz, when the auxiliary qubit is biased to its maximum frequency, the sweet spot (see Table~\ref{table:paramp} for a summary of the parameters). We use qubit spectroscopy to find the transition frequencies $\omega_{ge}/2\pi=5.758$~GHz and $\omega_{ef}/2\pi=5.455$~GHz. Time-resolved resonator spectroscopy when preparing the first (second) excited state of the auxiliary qubit allows for the extraction of the dispersive shift of that state imparted on the readout resonator $\chi^{(e)}/2\pi=-5$~MHz ($\chi^{(f)}/2\pi=-8$~MHz).
\begin{table}[!t]
\centering
 \begin{tabular}{l l r l}
 \hline
 Auxiliary & $g$-$e$ frequency, $\omega_{ge}/2\pi$ & 5.758 & GHz \\
 & $e$-$f$ frequency, $\omega_{ef}/2\pi$ & 5.455 & GHz \\
 & anharmonicity, $\alpha/2\pi$ & -303 & MHz \\
 & lifetime of $\ket{e}$, $T_1^{(e)}$ & $21$ & $\mu$s \\
 & lifetime of $\ket{f}$, $T_1^{(f)}$ & $7$ & $\mu$s \\
 & Ramsey dephasing time of $\ket{e}$, $T_2^{\star(e)}$ & $17$ & $\mu$s \\
 & Ramsey dephasing time of $\ket{f}$, $T_2^{\star(f)}$ & $8$ & $\mu$s \\
 & readout frequency, $\omega_\text{ro.}/2\pi$ & 6.664 & GHz \\
 & readout linewidth, $\kappa_\text{eff.}/2\pi$ & 25 & MHz \\
 & dispersive shift of $\ket{e}$, $\chi^{(e)}/2\pi$ & -5 & MHz \\
 & dispersive shift of $\ket{f}$, $\chi^{(f)}/2\pi$ & -8 & MHz \\
 Coupler & tunable coupler frequency, $\omega_\text{c}/2\pi$ & 4.7 & GHz \\
 Emitter & $0$-$1$ frequency, $\omega_{01}/2\pi$ & 5.896 & GHz \\
 & decay rate, $\kappa/2\pi$ & 1.95 & MHz \\
 \hline
 \end{tabular}
 \caption{\label{table:paramp}Measured device parameters.}
\end{table}

In a second step, we use the coherent scattering of a weak input tone by the emitter qubit into the transmission line to determine its transition frequency. Using a bias coil, we tune it to $\omega_{01}/2\pi=5.896$~GHz, such that both the SWAP frequency $\Delta_\text{e0g1}/2\pi=139$~MHz and the CNOT frequency $\Delta_\text{f0e1}/2\pi=441$~MHz are well separated from each other, far detuned from the anharmonicity $\alpha/2\pi=(\omega_{ef}-\omega_{ge})/2\pi=-303$~MHz, and within the bandwidth of the Arbitrary Waveform Generator (AWG) used to drive them.

We characterize the tunable coupler by driving it into a mixed state using a strong continuous drive on the charge line which weakly couples to it, and observing a dispersive shift on both the emitter and the auxiliary qubits. Tuned to $\omega_\text{c}/2\pi=4.7$~GHz, the tunable coupler imposes a coupling between the auxiliary and the emitter qubits that has the opposite sign as the coupling of the fixed coupler, such that the constant coupling $J_\text{DC}$ vanishes. We verify experimentally that $J_\text{DC}/2\pi<20$~kHz is reachable by performing time-resolved $T_1$ and Ramsey sequences at smaller auxiliary-emitter detunings, and observe that the lifetime and Ramsey dephasing times of the auxiliary $\ket{e}$ and $\ket{f}$ states are not limited by their Purcell coupling to the emitter at the frequencies chosen for the experiment. We extract the strength of the parametrically activated couplings $J_\text{AC}/2\pi\simeq 5$~MHz from the splitting of the emitter transition when parametrically tuned into resonance with a transition of the auxiliary qubit.

By comparing the Rabi-oscillation visibilities of the $e$-$f$ transition with and without a $R_y^{ge}(\pi)$-pulse~\cite{Jin2015b}, we measure a steady-state thermally excited population of the auxiliary qubit of $n_\text{exc.}=0.08$. We realize an unconditional reset of the thermally excited population of the auxiliary qubit~\cite{Magnard2018}, by preceding all measurements reported in this manuscript by a 2~$\mu$s long pulse at the SWAP frequency $\Delta_\text{e0g1}$. This equates the auxiliary system excited population to the steady-state population of the emitter qubit, which quickly thermalizes to the output transmission line thermal population through the engineered decay rate $\kappa$. After a reset pulse, we extract, in single-shot readout, as well as by comparing the Rabi-oscillations visibilities on the $e$-$f$ transition with and without a $R_y^{ge}(\pi)$-pulse, an upper limit of $n_\text{th}=0.003$ of the excited-state population. This corresponds to a 40~mK equivalent temperature of the transmission line.

\begin{figure}[!t]
\includegraphics[width=\columnwidth]{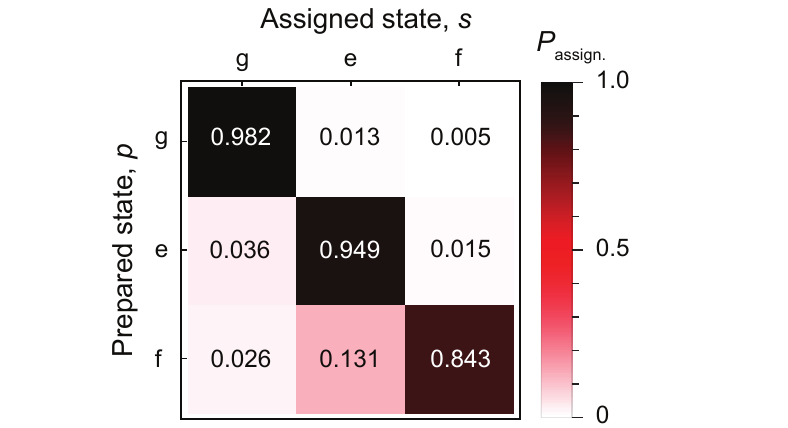}
\caption{\label{fig:assignmentmatrix}\textbf{Qutrit readout.} Assignment probability matrix $P_\text{assign.}(s|p)$ for the auxiliary system $A$ prepared in state $p$, and assigned to state $s$.}
\end{figure}

We perform three-level dispersive single-shot readout~\cite{Magnard2018} in 256~ns a fidelity characterized by the assignment probability matrix $P_\text{assign.}$ presented in Fig.~\ref{fig:assignmentmatrix}.

\section{Two-qubit gates}\label{SM:2qbgates}
We perform gates of the SWAP and CNOT families, which are represented with respect to the basis $B=\left\lbrace \ket{g0},\ket{g1},\ket{e0},\ket{e1} \right\rbrace$ by an arbitrary rotation angle $\theta$ and phase $\phi$ as
\begin{equation}
\text{SWAP}(\theta,\phi) =
\begin{pmatrix}
1  &  0  &  0  &  0  \\
0  &  \cos\theta/2  &  e^{i\phi}\sin\theta/2   &  0  \\
0  &  e^{-i\phi}\sin\theta/2  &  \cos\theta/2   &  0  \\
0  &  0  &  0  &  1 \\
\end{pmatrix}
\end{equation}
and
\begin{equation}
\text{CNOT}(\theta,\phi) =
\begin{pmatrix}
1  &  0  &  0  &  0  \\
0  &  1  &  0  &  0  \\
0  &  0  & \cos\theta/2  &  e^{i\phi}\sin\theta/2\\
0  &  0  &  e^{-i\phi}\sin\theta/2  &  \cos\theta/2 \\
\end{pmatrix}.
\end{equation}
In the work reported here, we always choose the phase $\phi=0$ by an appropriate choice of the phase of the radio-frequency pulses, and start a two-qubit gate with the emitter in the state $\ket{0}$.

\begin{figure}[!t]
\includegraphics[width=\columnwidth]{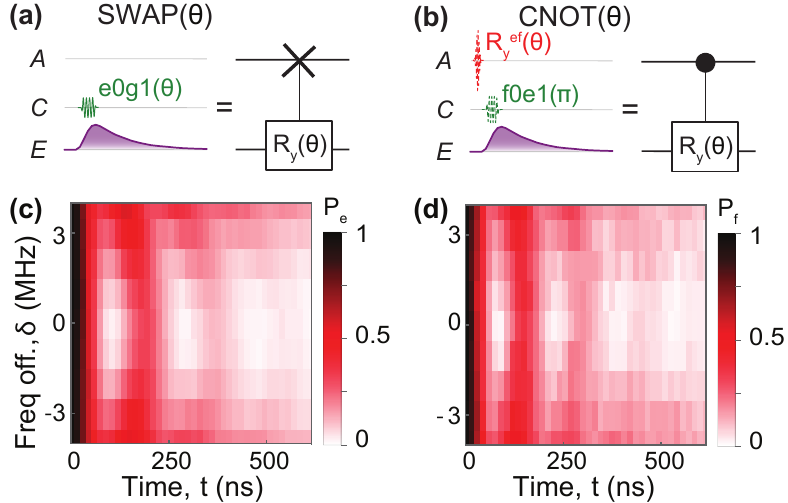}
\caption{\label{fig:chevrons}\textbf{SWAP and CNOT gate calibration.} (a-b) Pulses applied to the auxiliary qubit ($A$) and the coupler ($C$) generating excitations in the emitter qubit ($E$), for a gate of the SWAP (a) or CNOT (b) family. (c)[(d)] Measured $\ket{e}$-state ($\ket{f}$-state) population as a function of pulse length $t$ and pulse detuning $\delta$ from the sideband frequency $\Delta_\text{e0g1}$ ($\Delta_\text{f0e1}$) for the pulse scheme shown in (a) [(b)].}
\end{figure}

We perform SWAP-type gates by applying a radio-frequency pulse to the coupler flux line at the difference frequency $\Delta_\text{e0g1}/2\pi=(\omega_{01}-\omega_{ge})/2\pi= 139~$MHz between the two states in the first manifold of excitations, Fig.~\ref{fig:chevrons}(a). We calibrate the duration and exact frequency of this pulse by measuring the excited state population of the auxiliary qutrit after preparing the $\ket{e0}$ state, for a pulse frequency offset $\delta$ and a duration $t$, see Fig.~\ref{fig:chevrons}(c). We choose the duration of the pulse $t$ to achieve a targeted rotation angle $\theta$.
We implement the CNOT-type gates in an analogous fashion: in a first step, we realize the conditional creation of an excitation by applying a $R_y^\textrm{ef}(\theta)$-pulse on the second transition of the auxiliary qutrit. Then, we apply a microwave pulse to the coupler flux line at the difference frequency $\Delta_\textrm{f0e1}/2\pi=(\omega_{01}-\omega_{ef})/2\pi=441~$MHz between the two states in the second manifold of excitations, Fig.~\ref{fig:chevrons}(b). We experimentally calibrate the duration and exact frequency of this pulse by preparing the $\left| f0 \right\rangle$ state, and recording the population $P_f$ remaining in the second-excited level of the auxiliary qubit for a pulse frequency offset $\delta$ and duration $t$, see Fig.~\ref{fig:chevrons}(d).
By always performing a pulse with the duration corresponding to a full transfer of excitation, and selecting the desired angle $\theta$ by the rotation angle of the excitation-creating pulse $R_y^\textrm{ef}(\theta)$, we guarantee to have no population remaining in the non-computational state $\left| f \right\rangle$ after the gate.

\begin{figure}[!t]
\includegraphics[width=\columnwidth]{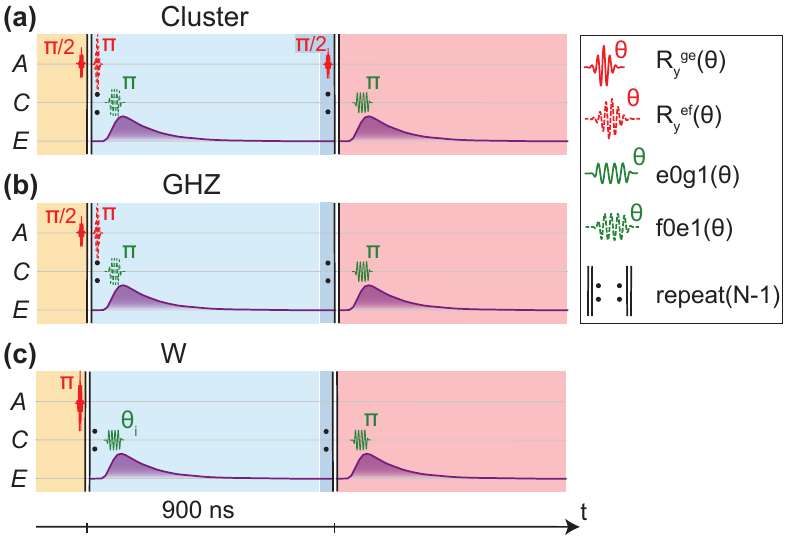}
\caption{\label{fig:pulses-applied-protocol}\textbf{Emission sequences.} Pulses applied to the auxiliary qubit ($A$) and the coupler ($C$) generating excitations in the emitter qubit ($E$), in order to prepare (a) the cluster state, (b) the GHZ state, and (c) the W state. Time axis (horizontal) is to scale, with photons emitted every 900~ns.}
\end{figure}

We implement the protocol of Fig.~\ref{fig:circuit_and_timetraces}(a-b) through the application of microwave pulses to the auxiliary qubit charge line and the coupler flux line, according to the timing diagram shown in Fig.~\ref{fig:pulses-applied-protocol}, where we plot the duration of the pulses to scale.
We emitted all photons with the same time delay of $T=900$~ns, chosen to be approximately one order of magnitude longer than the decay time of the emitter $\kappa^{-1}\simeq 80$~ns, such that the photonic time-bins are well separated. Further improvements in the fidelity and entanglement length could be achieved in the future by reducing the repetition time $T$ and thereby finding the best trade-off between coherent and incoherent errors. A choice of larger decay rate $\kappa$ of the emitter would also allow the repetition time to be decreased.

\section{Complete photonic tomography}\label{SM:tomography}
The procedure for complete tomography of photonic states with up to $N=4$ modes is based on the $R\rho R$ method described in Ref.~\cite{Eichler2012}, and detailed below. The signal amplitude $a_\text{out}$ emitted from the sample passes through an amplification chain whose first element is a JPA with 18~dB of phase-preserving gain. After further amplification and filtering, it is down-converted at room temperature to 250~MHz using a local oscillator, and digitized at 1~GS/s by a analog-to-digital converter (ADC) passing the data on to a field-programmable gate array (FPGA).

\begin{figure}[!t]
\includegraphics[width=\columnwidth]{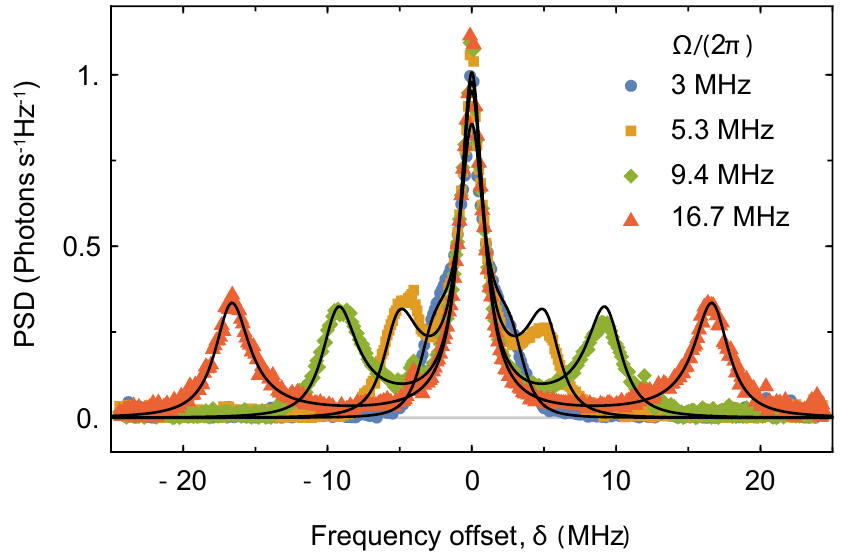}
\caption{\label{fig:MollowTriplet}\textbf{Absolute power calibration.} Measured power spectral density (PSD) of the inelastic scattering of a coherent tone resonant with the emitter qubit
(symbols) for various drive rates $\Omega$. Solid lines represent a global fit to the data.}
\end{figure}

We use the resonant photon-blockade of the emitter qubit under strong drive~\cite{Lang2011} to realize a calibrated power source and quantify the root-mean-square (RMS) voltage at the ADC input corresponding to one photon being emitted by our sample. This is performed by continuously, resonantly, driving the emitter qubit at rate $\Omega > \kappa$ and recording the inelastically emitted radiation, see Fig.~\ref{fig:MollowTriplet}. The nonlinear power spectral density (PSD), showing satellite peaks at detunings $\delta\simeq \pm \Omega$, and saturating to 1~photon~s~$^{-1}$~Hz$^{-1}$, is globally fitted as a Mollow triplet. This fit serves as a calibration for the emitted power $P=n_q \kappa \hbar \omega_{01}$. Here, $n_q\simeq 1/2$ is the steady-state average excited-state population of the emitter qubit under a large drive rate $\Omega$.

On the FPGA, we integrate the amplified output signal $s_{i,\text{out}}(t)$ in each time bin with a mode-matched filter $w_i(t)$ satisfying $\int \left| w_i(t) \right|^2 \text{d}t =1$ to yield the complex amplitude $S_i=\int \text{d}t \, w_i(t)s_{i,\text{out}}(t) = I_i + i Q_i = a + h^\dagger$. Here, $a$ is the mode of interest, and $h$ the added noise in the detection chain. We collect a quadrature pair $\left\lbrace I_i,Q_i \right\rbrace$ per photonic mode $P_i$. We record two $2N$-dimensional histograms of the measured distributions of the quadratures, the first one ($H_\text{on}$) with the signal mode $a$ in the emitted state $\ket{\psi}$ to be characterized, and added noise of the detection chain, the second one ($H_\text{off}$) with the signal mode $a$ prepared in the vacuum state, i.e.~recording noise added by the detection chain only. Memory constraints on the acquisition device limit the total number of histogram bins to $2^{24}$, resulting in a resolution of $2^{24/2N}$ bins for each quadrature, which currently limits us to perform full tomography for up to $N=4$ modes.
Since the Mollow triplet based calibration described above provides us with an absolute scale for the histogram axes, we quantify the efficiency of the detection chain $\eta = (1+n_\text{noise})^{-1} \approx 0.29$, corresponding to $n_\text{noise} \approx 2.5$ added noise photons, by computing the average photon number based in $H_\text{off}$. This efficiency is a bit lower than state of the art, due to the low gain (18~dB) chosen to preserve the linearity of the chain, such that the amplified vacuum noise by the JPA does not fully overcome the noise added later in the chain.

\begin{figure}[!t]
\includegraphics[width=\columnwidth]{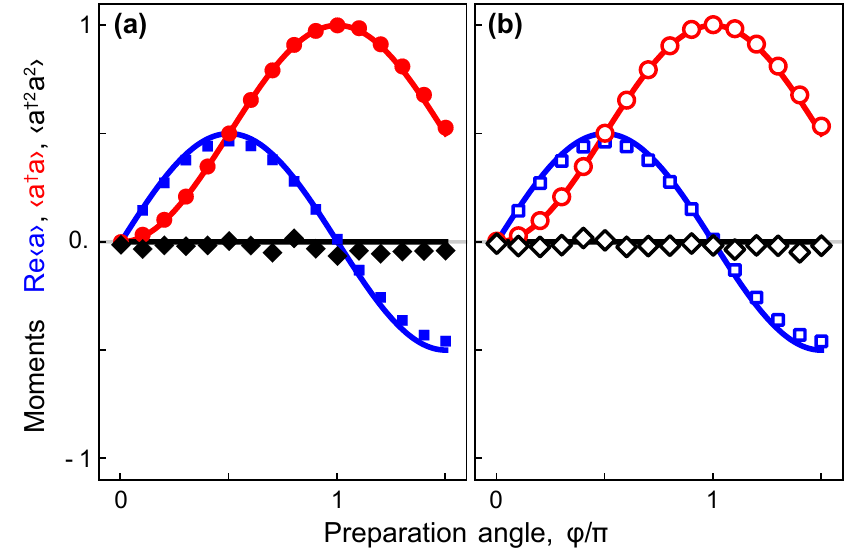}
\caption{\label{fig:SinglePhoton}\textbf{States in the single photon manifold.} Measured moments $\text{Re}\langle a\rangle$ (blue squares), $\langle a^\dagger a \rangle$ (red circles), and $\langle (a^\dagger)^2 a^2 \rangle$ (black diamonds), for superpositions $\ket{\psi}=\sin (\varphi/2)\ket{0}+\cos (\varphi/2)\ket{1}$ of vacuum and single photon states. The states are generated from the first manifold of the auxiliary qubit (a, filled markers) using SWAP gate, or the second manifold (b, empty markers) using a CNOT gate. Lines indicate ideal expectation values. Imaginary parts, not shown, are all below 0.05. Statistical error bars are smaller than the markers.}
\end{figure}

We verify the single photon character by measuring anti-bunching performing single mode tomography~\cite{Eichler2011}. We prepare a quantum superposition of vacuum and a single photon in the Fock basis $\ket{\psi}=\sin (\varphi/2)\ket{0}+\cos (\varphi/2)\ket{1}$ in two different ways. The first is achieved by preparing the equivalent superposition between the ground and first excited state of the auxiliary qubit $\ket{\psi_1}=\sin (\varphi/2)\ket{g}+\cos (\varphi/2)\ket{e}$ and performing a SWAP gate. The second consists of the prepared superposition in the second manifold $\ket{\psi_2}=\sin (\varphi/2)\ket{e}+\cos (\varphi/2)\ket{f}$, followed by a $\ket{f0}-\ket{e1}$ transition. We evaluate the expectation values of the moments $\left\langle (a^\dagger)^n a^m \right\rangle$ up to $n,m < 3$ versus the preparation angle $\varphi$, see Fig.~\ref{fig:SinglePhoton} for the real part of the amplitude $\text{Re}\langle a \rangle$, the power $\langle a^\dagger a \rangle$, and the second-order correlator $\langle (a^\dagger)^2 a^2 \rangle$.
We verify that our gates lead to states in the single photon manifold, as the $\left\langle (a^\dagger)^2 a^2 \right\rangle$ moments measured are all close to zero, with extreme values $-0.07$ and $+0.06$, corresponding to a vanishing second-order correlation function $g^{(2)}(0)$ at zero time delay. The emitted power $\left\langle a^\dagger a \right\rangle$ follows the expected $\sin^2(\varphi)$ dependence well in both cases. The coherence, characterized by the first order moment $\left\langle a\right\rangle$, is slightly reduced from the ideal value in both cases in agreement with the Ramsey-dephasing times of the $\ket{e}$ and $\ket{f}$ levels.

For photonic states with multiple modes, we start by verifying that the second-order correlation function $\langle (a^\dagger)^2 a^2 \rangle$ vanishes in each individual mode. We reconstruct the density matrix of the noise mode $\rho_h$ by defining positive operator-valued measures (POVMs) $\Pi_j=\Pi_{\otimes_i S_i} = \bigotimes_i \Pi_{S_i}$, where $\Pi_{S_i} = \pi^{-1} \ket{\alpha=S_i}\bra{\alpha=S_i}$. We create one POVM per histogram bin, using the absolute scale found in the Mollow triplet fit Fig.~\ref{fig:MollowTriplet}. We then iteratively converge towards $\rho_h$ by starting in a maximally mixed state $\rho_0=\mathbbm{1}/d$ and updating the estimate according to
\begin{equation}
\rho_{k+1} = N G^{-1} R(\rho_k) \rho_k R(\rho_k) G^{-1},
\end{equation}
where $R(\rho)=\sum_j H_\text{off}(j) \Pi_j [\text{Tr}(\rho \Pi_j)]^{-1}$, $G=\sum_j \Pi_j$, and $N$ is a renormalization constant. This procedure guarantees convergence to the most likely physical density matrix $\rho_h$ of the noise mode. Experimentally, we find very good agreement between the most likely matrix found via the iterative method and a thermal state $\rho_\text{th.}=\sum_{n=0}^\infty n_\text{noise}^n/(1+n_\text{noise})^{(n+1)} \ket{n}\bra{n}$ with $n_\textrm{noise}\approx 2.5$ noise photons. Nevertheless, we use the experimental noise mode $\rho_h$ in the following.
We then apply the iterative method a second time, with the knowledge of the noise mode enabling the creation of POVMs that are displaced noise modes $\Pi_{S_i} = \pi^{-1} D_{\alpha=S_i}\rho_h D^\dagger_{\alpha=S_i}$, with the displacement operator $D$, and using the histogram $H_\text{on}$ instead of $H_\text{off}$ to construct the update operator $R$. We reconstruct with this procedure the most likely density matrix $\rho$ of the emitted state $\ket{\psi}$ under the physicality constraints, that is, the matrix is hermitian, positive semi-definite, and its trace is equal to 1.

\begin{figure}[!t]
\includegraphics[width=\columnwidth]{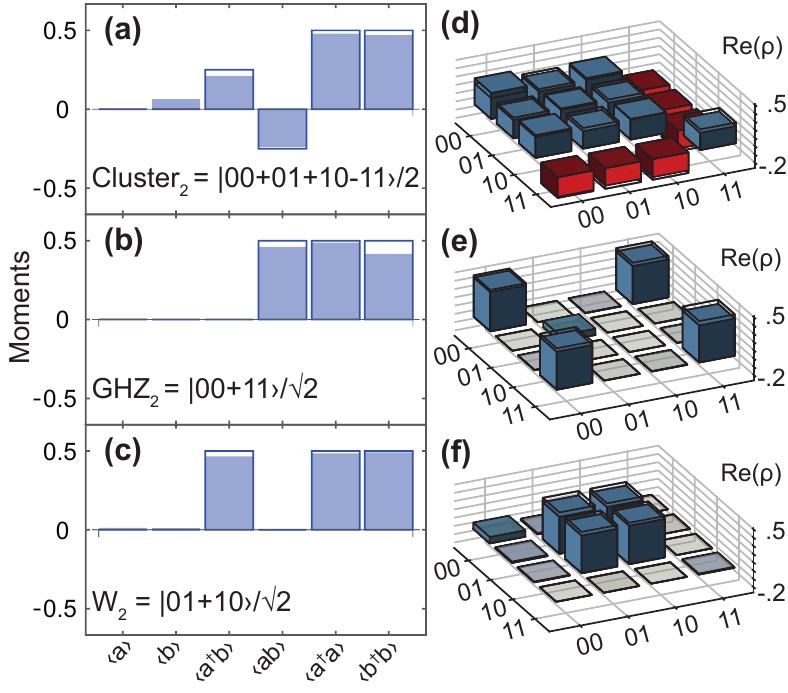}
\caption{\label{fig:Bells}\textbf{Two-mode states with Bell-type correlations.} Two-photon state tomography for (a,d) the cluster state, (b,e) the GHZ state (Bell state $\ket{\Phi^+}$) and (c,f) the W state (Bell state $\ket{\Psi^+}$). (a-c) Real part of the moments listed (blue bars), together with ideal values (solid blue wireframes). The operator $a$ ($b$) refers to the first (second) time-bin defined photonic state. (d-f) Real part of the density matrices $\text{Re}(\rho)$ (bars) and ideal values $\text{Re}(\rho_\text{ideal})$ (black wireframes). All imaginary parts of moments and density matrix entries are below 0.02.}
\end{figure}

We reconstruct the two-mode states, Bell states up to local rotations, emitted using the protocols described above, see Fig.~\ref{fig:Bells}. We observe, as expected, that each mode shows almost zero coherence when tracing out the other one (first order moments $\langle a \rangle$ or $\langle b \rangle$ close to zero). The cross-correlators of the form $\langle ab \rangle$ or $\langle a^\dagger b \rangle$ are non-zero, indicating entanglement between the two modes. We reconstruct density matrices with fidelities $F_{\text{Cluster}_2}=0.94$, $F_{\mathrm{GHZ}_2}=0.92$, and $F_{\mathrm{W}_2}=0.91$, and negativities $\mathcal{N}_{\text{Cluster}_2}=0.44$, $\mathcal{N}_{\text{GHZ}_2}=0.42$, and $\mathcal{N}_{\text{W}_2}=0.41$ (ideal value is 0.5) as a witness of entanglement.

Density matrices for the states with three and four photonic qubits are discussed in the main text.

\section{Process maps}\label{SM:maps}
As motivated in the main text we estimate the density matrices for states with $N>4$ within the class of matrix products density operators with bond dimension $d=2$ by experimentally reconstructing the process map $\chi^{(l)}$ for individual emission processes and by calculating $\rho = (\Pi_{l=N}^1 \chi^{(l)})\rho_0$, where $\rho_0$ is the density matrix obtained after the initialization pulse in Fig.~\ref{fig:circuit_and_timetraces}. We further assume that the maps $\chi^{(l)}$ for nominally identical processes are independent of $l$, which assumes history-independent operations. This assumption is justified as we always perform the same operation, and after one photon is emitted, it does not interact with the auxiliary system anymore. In order to estimate $\rho$ for the cluster and GHZ states we therefore need to characterize four process maps in total, for the CNOT, the Hadamard followed by a CNOT, the SWAP, and the Hadamard followed by a SWAP. Each process matrix $\chi^{(l)}$ maps an input state of the auxiliary qubit to an output state of the joint system consisting of the auxiliary qubit and the emitted photonic qubit $\chi: \rho_A^{(\text{pre})} \rightarrow \rho_{A,P}^{(\text{post})}$. In the Pauli basis, we write the auxiliary qubit input state $\rho_A^{(\text{pre})} = \sum_k \rho_k^{(\text{pre})} \sigma_k$ and the output joint system
$
\rho_{A,P}^{(\text{post})} = \sum_{i,j} \rho_{i,j}^{(\text{post})} \sigma_i \otimes \sigma_j = \sum_{i,j} \left(\sum_k \chi_{i,j}^k \rho_k^{(\text{pre})}\right) \sigma_i \otimes \sigma_j.
$

\begin{figure}[!t]
\includegraphics[width=\columnwidth]{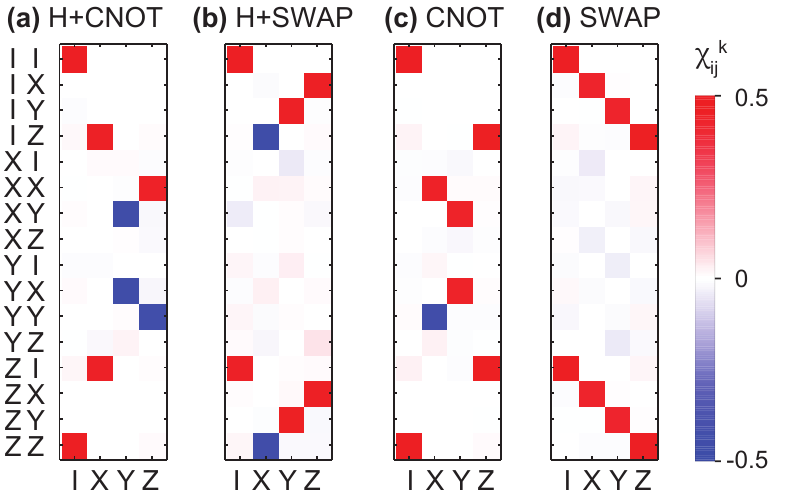}
\caption{\label{fig:process-maps}\textbf{Measured process maps of repeatable operations.} Process maps $\chi$ in the Pauli basis. (a,b) Operations used for the cluster state preparation: H+CNOT (a), H+SWAP (b). (c,d) Operations used for the GHZ state preparation: CNOT (c), SWAP (d).}
\end{figure}
We characterize a process map by recording correlations between the auxiliary system and the emitted photonic qubit. For each prepared cardinal state of the auxiliary qubit, $\left\lbrace \ket{g}, \ket{e}, \ket{g+e}, \ket{g-e}, \ket{g+i e}, \ket{g-ie} \right\rbrace$ (4 cardinal states would have been sufficient, but we use redundancy to reduce the impact of experimental errors), we record 3 state-conditioned 2D-histograms $H\left(I,Q,s \right)$~\cite{Eichler2012b}, where $I$ and $Q$ are mode-matched integrated quadratures of the photonic mode as in the previous section, and $s$ is the state the auxiliary qutrit was assigned to, in single-shot dispersive readout~\cite{Walter2017}, directly on the FPGA. The 3 histograms correspond to the $X,Y,Z$-bases of measurement for the auxiliary qubit. We correct for finite readout assignment fidelity by inverting the assignment probability matrix presented in Fig.~\ref{fig:assignmentmatrix}, and by processing the histograms $\tilde H\left(I,Q,s \right) = \sum_{s'} P_\text{assign.}^{-1}({s'}|s) H(I,Q,{s'}) $.

We verify that the $\ket{f}$-state population is below 0.01 in all cases, allowing us to treat the auxiliary system as a qubit to very good approximation, and reconstruct the joint auxiliary and photonic qubits output density matrix for each of the six input states~\cite{Eichler2012b}. These matrices yield an overdetermined set of equations for the process map, of the form $\rho_{i,j}^{(\text{post})} = \sum_k \chi_{i,j}^k \rho_k^{(\text{pre})}$, which we solve for each fixed value of $i$ and $j$ using least squares, finding $\chi_\text{LS}$. We impose physicality in a last step by ensuring that the map is trace preserving, and eigenvalues of the corresponding $8\times 8$ Choi matrix $C^\chi$ are non-negative following Ref.~\cite{Schwartz2016}. We obtain the least squares Choi matrix $C^\chi_\text{LS} = \sum_{l,m} e_{l,m} \otimes \left( \sum_{i,j} \sum_k \chi_{i,j}^k \Lambda_{lm,k} \sigma_i \otimes \sigma_j \right)$ from the process map $\chi_\text{LS}$. Here, $e_{l,m}$ is a $2\times 2$-matrix with value 1 in the $l,m$-entry and 0 otherwise, and
\begin{equation}
\Lambda_{lm,k}=\frac{1}{2} \begin{pmatrix}
1  &  0  &  0  &  1  \\
0  &  1 &  i   &  0  \\
0  &  1  &  -i   &  0  \\
1  &  0  &  0  &  -1 \\
\end{pmatrix}
\end{equation}
is the transformation matrix from the Pauli basis to the computational basis. Choi's theorem guarantees that the process map is completely positive and trace preserving if and only if its Choi matrix is a positive matrix satisfying the conditions $\text{Tr}[C^\chi \sigma_0 \otimes \sigma_0 \otimes \sigma_0] = 2$ and $\text{Tr}[C^\chi \sigma_i \otimes \sigma_0 \otimes \sigma_0] = 0 \, \, \forall i\neq 0$.
We find numerically the matrix $C^\chi$ which minimizes the distance to $C^\chi_\text{LS}$ (measured as the norm of the difference of the two matrices) under these constraints of physicality. The corresponding process maps, shown in Fig.~\ref{fig:process-maps}, have real entries in a Pauli transfer matrix representation, and are close to the ideally expected ones with fidelities $F_\text{CNOT}=0.924$, $F_\text{SWAP}=0.918$, $F_\text{H+CNOT}=0.923$, and $F_\text{H+SWAP}=0.928$. The fidelity is defined as $F=\text{Tr}\left( \sqrt{\sqrt{C^\chi} C^\chi_\text{ideal} \sqrt{C^\chi}} \right)^2$, by comparing the experimental Choi matrix $C^\chi$ to the ideal one $C^\chi_\text{ideal}$.


%

\end{document}